\documentclass[fleqn,twoside]{article}
\usepackage{espcrc2}
\newif\ifpdf
\ifx\pdfoutput\undefined\pdffalse\else\pdfoutput=1\pdftrue\fi
\ifpdf
  \usepackage[backref,citecolor=blue]{hyperref}
  \let\myhref=\href\def\href#1#2{\penalty0\myhref{#1}{\tt #2}}%
\else%
  \def\href#1#2{{\penalty0\tt #2}}
\fi
\def\epspdffile#1{\leavevmode\ifpdf\epsffile{#1.pdf}%
  \else\epsffile{#1.eps}\fi}
\usepackage{epsfig}
\usepackage{amsfonts}			% blackboard bold stuff
\usepackage[OT2,OT1]{fontenc}		% define font encoding schemes
\newcommand\cyr{%
  \renewcommand\rmdefault{wncyr}%	% roman
  \renewcommand\bfdefault{b}%		% bold
  \renewcommand\encodingdefault{OT2}%
  \normalfont\selectfont}%
\DeclareTextFontCommand{\textcyr}{\cyr}
\DeclareTextSymbol{\cprime}{OT2}{'176}	% Cyrillic '
\def\Zolotarev{\textcyr{Zolotarev}}
\def\Chebyshev{\textcyr{Chebyshev}}
\def\Remez{\textcyr{Remez}}
\def\Jacobi{Jacobi}
\def\Krylov{\textcyr{Krylov}}
\def\Liapunov{\textcyr{Lyapunov}}
\def\Luscher{L\"uscher}
\def\HMDR{\(R\)}			% HMD R algorithm
\def\SU#1{\mathop{\rm SU}(#1)}		% SU(n) Lie group
\def\su#1{\mathop{\rm su}(#1)}		% su(n) Lie algebra = A(n-1)
\def\asqtad{{\scshape asqtad}}		% a^2 tadpole-improved
\def\qcdtm{\(\mbox{QCD}^{\mbox{\tiny TM}}\)} % Twisted mass QCD
\def\M{{\cal M}}			% Fermion kernel
\def\D{{\cal D}}			% Diagonal matrix
\def\t{\tau}				% MD time
\def\dt{\delta\t}			% MD step size
\def\H{H}				% Hamiltonian
\def\dH{\delta\H}			% Change in Hamiltonian
\def\refsec#1{\S\ref{#1}}		% Cross-reference to section
\def\reffig#1{Figure~\ref{#1}}		% Cross-reference to figure
\def\defn{\equiv}			% definition
\def\tD{\mbox{D}\kern-0.65em\raise0.15ex\hbox{/}\kern0.15em} % D slash
\def\sD{\mbox{\scriptsize D}\kern-0.5em\raise0.15ex\hbox{\scriptsize/}}
\def\ssD{\mbox{\tiny D}\kern-0.42em\raise0.15ex\hbox{\tiny/}}
\def\Dslash{{\mathchoice{\tD}{\tD}{\sD}{\ssD}}}
\def\DWslash{\Dslash_W}			% D_W slash
\def\dslash{\hbox{\(\partial\)}\kern-0.5em\raise0.15ex\hbox{/}} % d slash
\def\rloc{{r_{\mbox{\tiny loc}}}}	% localisation radius
\def\opt{{\mbox{\tiny opt}}}		% "opt" subscript
\def\pf{{\mbox{\tiny pf}}}		% "pf" subscript
\def\min{\mathop{\rm min}}		% minimum
\def\max{\mathop{\rm max}}		% maximum
\def\det{\mathop{\rm det}}		% determinant
\def\tr{\mathop{\rm tr}}		% trace
\def\ln{\mathop{\rm ln}}		% natural logarithm
\def\sgn{\mathop{\rm sgn}}		% signum
\def\Re{\mathop{\rm Re}}		% Real Part
\def\rational#1#2{{\mathchoice{\textstyle{#1\over#2}}%
  {\scriptstyle{#1\over#2}}{\scriptscriptstyle{#1\!/\!#2}}{#1\!/\!#2}}}
\def\half{\rational12}                      % One half
\begin{document}

\title{Algorithms for Lattice QCD with Dynamical Fermions}

\author{A. D. Kennedy\address{School of Physics, University of Edinburgh,
 King's Buildings, Edinburgh, EH9 3JZ, United Kingdom}}

\begin{abstract}
  {\noindent We consider recent progress in algorithms for generating gauge
  field configurations that include the dynamical effects of light fermions. We
  survey what has been achieved in recent state-of-the-art computations, and
  examine the trade-offs between performance and control of systematic
  errors. We briefly review the use of polynomial and rational approximations
  in Hybrid Monte Carlo algorithms, and some of the theory of on-shell chiral
  fermions on the lattice. This provides a theoretical framework within which
  we compare algorithmic alternatives for their implementation; and again we
  examine the trade-offs between speed and error control.\parfillskip=0pt\par}
\end{abstract}

\maketitle

\section{Introduction}

The aim of this review is to provide a snapshot of the present status of
dynamical fermion simulations, both as regards their performance and the
possible sources of systematic errors that are not fully under control, as well
as an introduction to some of the algorithmic ideas that are currently being
investigated. As the goal is to be pedagogical rather than exhaustive we do not
attempt to summarise all of the algorithmic talks that were presented at this
conference.

In \refsec{sec:battlefield} we consider the present state of large-scale
dynamical fermion computations, comparing their costs and their control of
systematic errors. This leads us to consider the issue of the locality of using
fractional powers of the fermion determinant in \refsec{sec:locality}. In
\refsec{sec:fat-links} we consider various algorithms for introducing ``fat
links'' in a differentiable way. In \refsec{sec:SAP} we review a new algorithm
which allows a domain-decomposition approach to be used for lattice field
theories. In \refsec{sec:fractional-multiplets} we outline the theory behind
some of the algorithms that are being used to include fractional powers of the
fermion determinant; the same techniques may be applied to evade instabilities
in numerical integrators (\refsec{sec:instability}), as can some other clever
tricks (\refsec{sec:integrator-tricks}). Finally, we consider in some detail
algorithms for including the dynamical effects of on-shell chiral fermions in
\refsec{sec:chiral}, and in particular emphasise that all the methods used may
be considered to be different approximations to the Neuberger operator.

\section[Numerical Simulations under "Battlefield Conditions"]{Numerical
Simulations under\\``Battlefield Conditions''} \label{sec:battlefield}

Before comparing the performance and efficacy of the various dynamical fermion
algorithms currently being used in large-scale numerical computations there
some caveats that must be emphasised.

Since such computations are extremely expensive they have only been carried out
at a small set of parameter values, and thus we cannot reliably compare their
results or behaviour in either the continuum or thermodynamic (infinite volume)
limit. Indeed, there are usually only results for two or fewer different
lattice spacings.

For the same reason it is not possible to compare the performance of the
algorithms at the same physical quark masses; for this we would need to
interpolate between runs with different dynamical (not valence) quark masses.

Some computations have been carried out with two quark flavours and others with
three, and again we are forced to ignore these differences due to lack of
data. Since the cost of adding in a third heavy dynamical quark is usually
small compared to the cost of the two lightest quarks this probably is not too
important within the large error bars.

It is very hard to make reliable estimates of the autocorrelation times
involved, and thus to estimate the cost of generating statistically independent
configurations from an equilibrium distribution. Not only do results depends on
the operator whose integrated autocorrelations are measured, but also on the
physics of the system, such as the presence of nearby phase transitions.

Computations have been carried out on different lattice sizes, so we obviously
need to compare the ``cost per unit volume.'' It would be wrong to assume
na{\"\i}vely that the cost of all algorithms scale linearly with the number of
lattice sites \(V/a^4\) at fixed lattice spacing \(a\): while the {\HMDR}
algorithm (\refsec{sec:hmdr}) cost scales as \(V\) for fixed
(volume-independent) \(O(\dt^2)\) errors in the parameters of the effective
action being simulated, the cost of the Hybrid Monte Carlo (HMC)
algorithm\footnote{This is because it is necessary to scale the molecular
dynamics (MD) step size \(\dt\propto V^{\rational14}\) in free field HMC for a
constant acceptance rate. We shall see later (\refsec{sec:instability}) that
for HMC computations involving light fermions the step size \(\dt\) is
constrained by instability in the integrator rather than by the bulk behaviour
of the HMC Hamiltonian, so a \(V\) dependence might be more appropriate for HMC
too.}  grows as \(V^{\rational54}\).  Fortunately the effects of such different
volume extrapolations are small compared to the errors in the cost estimates
anyhow.

\begin{figure}[th]
  \epsfxsize=0.45\textwidth
  \centerline{\epspdffile{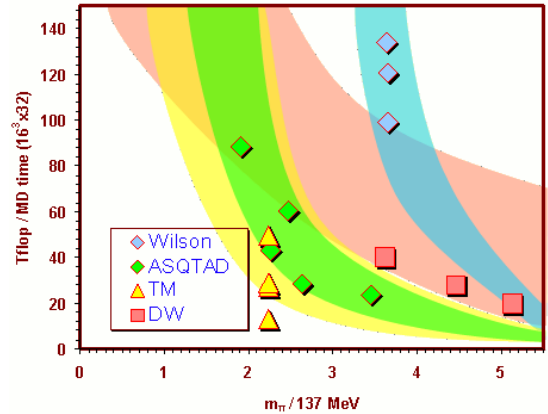}}
  \vskip-6ex
  \caption[Algorithm cost]{Comparison of cost of various algorithms. The Wilson
  fermion data is taken from \cite{Namekawa:2004bi,Ukawa:2002pc}, the {\asqtad}
  data from \cite{Aubin:2004wf,Gottlieb:2004}, the {\qcdtm} data from
  \cite{Jansen:2004}, and the domain wall data from \cite{Mawhinney:2004}. The
  scatter of the points gives a rough indication of the uncertainties in the
  cost estimates. For the {\qcdtm} fermions there are number of points at the
  same value for the \(\pi\) mass, because the system shows a first-order phase
  transition with different costs in each phase~\cite{Farchioni:2004us}. The
  bands are a subjective extrapolation of the costs.}
  \label{fig:algorithm-cost}
\end{figure}

\reffig{fig:algorithm-cost} shows the approximate cost of evolving a
\(16^3\times32\) lattice through one unit of MD time as a function of the
dynamical \(\pi\) mass achieved. Four types of fermion action have been used
recently for large-scale computations. \emph{Wilson/Clover} fermions seem to
increase in cost as the \(\pi\) mass is reduced much more rapidly than the
other formulations; indeed, there are runs at smaller \(\pi\) mass which are so
expensive that they lie significantly above the graph. It may be conjectured
that this is because the Dirac spectrum not bounded below, which is not the
case for \emph{KS/Staggered} fermions (where recent data from the MILC
collaboration use the {\asqtad} improved action), \emph{twisted mass} (\qcdtm)
fermions, or \emph{domain wall} (GW/Overlap) fermions. The question of which
formulation on-shell chiral fermions is best, or indeed whether such
formulations are to be preferred at all, depends on how much chiral symmetry is
required. At present dynamical GW fermions seem to be about 10--100 times more
expensive than the corresponding {\asqtad} or {\qcdtm} runs at comparable
masses.

The figure also indicates subjective extrapolations for the costs. In
particular it seems reasonable that the cost for dynamical domain wall
fermions, or any other formulation exhibiting exact on-shell lattice chiral
symmetry, should eventually become cheaper for light enough quarks because the
chiral limit for them does not require taking the continuum limit.\footnote{Of
course we need to be close enough to the continuum limit in all cases to
extract real-world physics.}

\section{Locality} \label{sec:locality}

There has been much debate recently about whether dynamical fermion
formulations that weight configurations with a non-integral power of the
fermion determinant correspond to local quantum field theories (QFTs) or not,
and how significant locality is. For example, staggered quarks appear in
multiples of four ``tastes;'' when the staggered fermion fields are integrated
over we are left with a fermion determinant \(\det\M\) in the functional
integral. This is replaced by \((\det\M)^\half\) to obtain a corresponding two
taste theory, and gauge configurations may be generated with this weight using
inexact (\HMDR)~(\refsec{sec:hmdr}) or exact (PHMC or RHMC)
algorithms~(\refsec{sec:RHMC}).

If a QFT is local then we are guaranteed that it has the cluster decomposition
property, and that within the context of renormalized perturbation theory it
satisfies the familiar power counting rules, exhibits universality, and is
amenable to a systematic improvement procedure. On the other hand, if it is not
local then there is little we can say about these properties other than that we
we have no \emph{a priori} reason to expect them to hold. In particular, if we
have a non-local lattice theory then we have no good reason to invoke
power-counting arguments to justify taking the na{\"\i}ve continuum limit, or
to expect the lattice theory to be described by continuum perturbation theory
however small the lattice spacing.

The fact that a formulation is not manifestly local does not logically imply
that it is not local. After all, it is easy to transform a manifestly local
theory into an equivalent but non-manifestly-local form: we use this freedom to
replace fermion fields with a non-manifestly-local fermion determinant, or an
equally non-manifestly-local pseudofermion representation. For the case of two
tastes of staggered fermions there may be a local fermion kernel \(\M'\) such
that \(\det\M' = \sqrt{\det\M}\), However, in general a non-manifestly local
theory has no reason to be equivalent to a local one.

Even if there was a local action corresponding to taking fractional powers of
the fermion determinant in the functional integral, we are still required to
use this local action to measure fermionic quantities. For the case of
staggered quarks this means one must use the hypothetical \(\M'\) rather than
\(\M\) for valence measurements. We should not expect that measuring operators
corresponding to a local four taste valence action \(\M\) on configurations
generated with \(\sqrt{\det\M}\) to lead to consistent results. Not only might
there be unknown renormalisations of the parameters between the sea and valence
actions (e.g., what is the justification for using the same numerical value for
the quark masses?) but the degrees of freedom are not even the same. Unless we
know \(\M'\) explicitly or are only interested in measuring purely gluonic
observables we are forced into having a ``mixed action,'' that is different sea
and valence actions, and such a situation suffers from the same problems, such
as violation of unitarity, as quenched QCD. Of course, one might hope that
these problems do not occur, or at least are smaller than for an arbitrary
mixed action situation, but these systematic errors are not quantitatively
under control.

\subsection[Is the Square Root of M Local?]{Is \(\sqrt\M\) Local?}

Some recent publications \cite{Bunk:2004br,Hart:2004sz} have studied the
question as to whether the operator \(\sqrt\M\) is in fact local. To make this
question meaningful we need to specify which square root is under
consideration. If the square root of the lattice operator is defined by taking
the square roots of its eigenvalues in a basis in which it is diagonal then the
sign of these roots is arbitrary. In the tests that have been carried out the
square root has been defined by a polynomial approximation, which corresponds
to choosing the positive square roots. Suppose we applied this procedure to the
Wilson Dirac operator \(\sqrt{\DWslash^\dagger\DWslash} = \sqrt{(\gamma_5
\DWslash)^2}\); we would find that the square root was non-local, because the
local square root (\(\gamma_5\DWslash\)) is not a positive operator.

It is important to stress that the fact that \(\sqrt\M\) appears to be
non-local does not prove that there is no corresponding local operator \(\M'\)
with the same determinant; on the other hand the only operator we know with
this determinant (and with staggered lattice symmetry) is \(\sqrt\M\) itself,
so it is worthwhile verifying that it is not (miraculously) local as has
sometimes been suggested as a possibility. After all, the Neuberger operator
appears to be local on sufficiently smooth gauge
configurations~\cite{Hernandez:1998et}, despite the fact that it is not
manifestly local. Hoping for the existence of \(\M'\), or even proving it, is
insufficient: we would need to use this operator for valence measurements in
order to have a manifestly consistent unitary theory.

Let us now briefly review how the locality of a lattice Dirac operator may be
measured numerically. We define a ``wavefunction'' by applying a lattice Dirac
operator to a point source, \(\psi(x) = a\Dslash(x,y) \delta(y)\). We then say
that the corresponding QFT is ultralocal if the wavefunction has ``compact
support,'' \(|\psi(x)| = 0\; \forall\|x-y\|>r\), and is local if the
wavefunction has ``fast decrease,'' \(\max_{\|x-y\|=r} |\psi(x)| \leq
Ce^{-r/\rloc}\), with \(\rloc\) being the localisation length. Note that in
both cases the wavefunction falls to zero at any non-zero physical distance in
the continuum limit, thus corresponding to a local continuum theory.

\begin{figure}[th]
  \epsfxsize=0.49\textwidth
  \centerline{\epspdffile{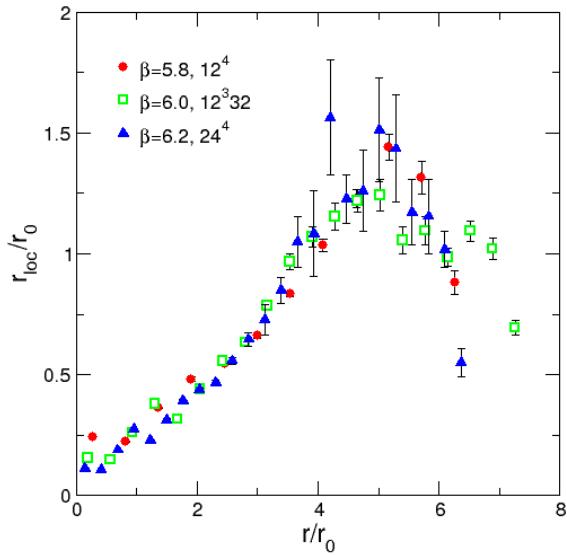}}
  \vskip-6ex
  \caption[Localisation]{Effective localisation length \(\rloc\) in physical
  units as a function of physical distance for the 2-taste {\asqtad}
  action, for a variety of different lattice spacings~\cite{Hart:2004sz}.} 
  \label{fig:localisation}
\end{figure}

Typical results for the effective localisation length \(a \left\{r_0 \ln[ \psi
(r)/\psi(r+a)]\right\}^{-1}\) are shown in \reffig{fig:localisation} as a
function of the physical distance (\(r_0\) is the Sommer scale). The effective
localisation length does not reach a plateau in the region where we may hope
that finite-size effects are under control, so we cannot extract a definite
estimate of \(\rloc\), on the other hand the evidence is that it must be
non-zero at a non-zero physical distance, independent of \(a\), so the operator
cannot be local in the continuum limit. Furthermore, the magnitude of the
localisation length is comparable with the inverse \(\pi\) mass, and thus use
of \(\sqrt\M\) would not lead to meaningful (valence) spectrum measurements.

The reader is also referred to interesting related work in the Schwinger
model~\cite{Durr:2003xs}.

\section{Fat Links} \label{sec:fat-links}

We shall now survey the topic of ``fat links.'' The idea of replacing gauge
link variables by some average of their neighbours is not new, and for a long
time has been used to construct operators that are good sources and sinks, in
the sense that they have a better overlap with ground state. The basic idea is
that instead of explicitly constructing a more complicated less local improved
operator we use the simplest operator with the correct symmetries (quantum
numbers) but taken as a function of the smeared or fattened link fields. The
fattening procedure is thus equivalent to building a physical size smeared
source. Of course, to calculate masses and so forth we need to measure temporal
correlation functions, and the sources and sinks for these should be localised
in time, so we typically want to use spatial link fattening only in this case.

Another use of fat links which is becoming important as dynamical fermion
computations become commonplace is that of smearing the action itself, in this
case by computing the original action as a function of the fattened link
fields. Here the goal is to suppress UV fluctuations in the gauge fields. This
may be considered as a means of constructing improved actions in the spirit of
renormalisation-group improved (or ``perfect'') actions, or systematic
improvement in powers of \(a\) (either perturbative or not).

There have recently been two main applications of fat links: improving gauge
actions so as to produce smoother gauge fields upon which quenched GW Dirac
operators are localised, and improving fermionic actions themselves. We shall
consider the latter here, as we are concerned with dynamical fermion
algorithms. The problem with using a highly improved action such as
HYP~\cite{Hasenfratz:2001hp} or perfect actions~\cite{Hasenfratz:1993sp} in
dynamical computations is that we either need to find non-MD-based algorithms
for generating the gauge configurations including the effect of the fermion
determinant, or we need to compute the MD force corresponding to the improved
actions.

\subsection{Molecular Dynamics with Fat Links}

The most successful fat actions, such as HYP, used in quenched computations
have been those where the fat links are somehow projected onto the group
manifold. Indeed, this seems to be a crucial ingredient for their success.
Unfortunately this projection would seem to be a non-differentiable operation
(i.e., the fat links are not a smooth function of the original thin links). The
first attempt to show that this was not necessarily so, and indeed that the MD
equations of motion for fat actions could be practicable, was using the FLIC
(Fat Link Improved Clover) action~\cite{Kamleh:2004xk}.

Let \(V\) be the usual unprojected APE-smeared link, then the usual iterative
projection \(V\to U\) where \(U\in\SU3\) maximizes \(\Re\tr UV^\dagger\) is not
differentiable; however by defining successively \(W \defn V/\sqrt{V
V^\dagger}\) and \(U \defn W/\root3\of{\det W}\) instead we obtain a mapping
that is almost differentiable. The ``almost'' is that differentiability fails
along the branch cut of the cube root, but this does not seem too important in
practice.

\begin{figure}[th]
  \epsfxsize=0.4\textwidth
  \centerline{\epspdffile{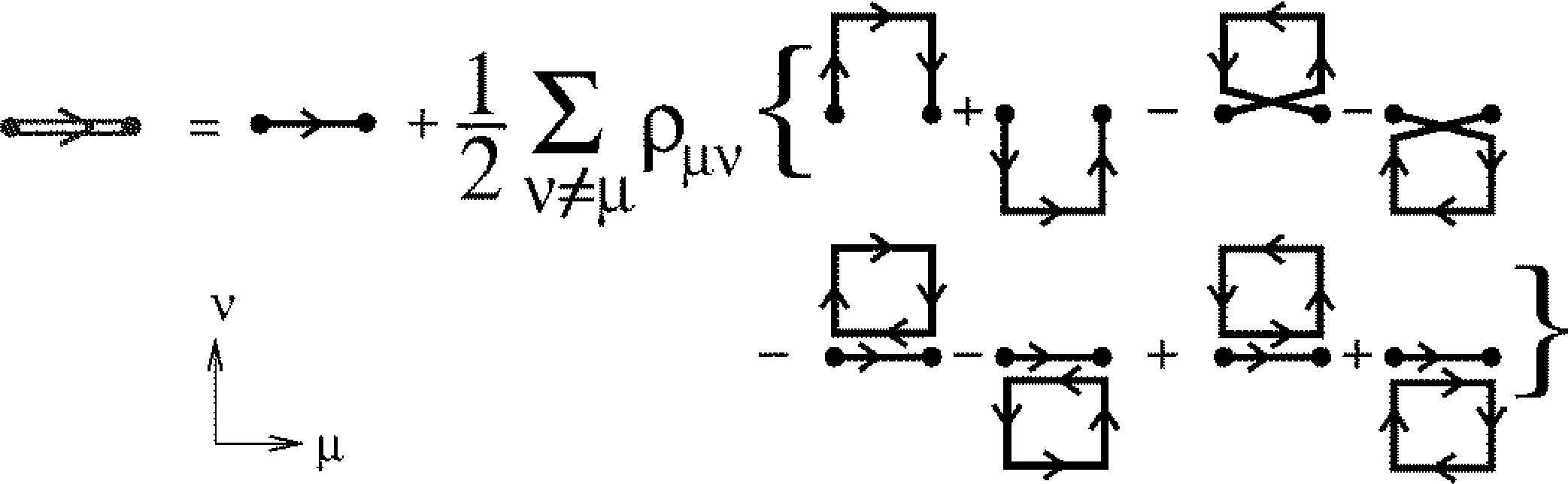}}
  \caption[Stout links]{Definition of stout links.}
  \label{fig:stout}
\end{figure}

Another ``projection'' procedure which is truly differentiable is that of
``stout links''~\cite{Morningstar:2003gk}. Let \(V\) be a suitably smeared link
and \(U_0\) the corresponding thin link, so that \(VU_0^\dagger\) is an
untraced sum of plaquettes. This may be ``projected'' onto \(\SU3\) by first
projecting it onto the \(\su3\) algebra and then exponentiating it into the
group, \(U = \exp T(VU_0^\dagger)\), where \(T\) means the traceless
antihermitian part. This is now differentiable, but it does not look too much
like projection (except when \(VU_0^\dagger\approx1\)). In quenched tests stout
links seem to be about as good as ordinary projected fat links, but require a
little more tuning.

These methods can be applied iteratively to produce differentiable links
of arbitrary obesity.

\section{Schwarz Alternating Procedure} \label{sec:SAP}

An interesting new algorithm based on the Schwarz Alternating Procedure (SAP)
has been introduced by {\Luscher} both as a preconditioner for solving the
lattice Dirac equation~\cite{Luscher:2003qa} and as dynamical fermion Monte
Carlo algorithm~\cite{Luscher:2003vf,Luscher:2004rx}. The SAP, introduced in
1870, was probably the first domain decomposition method for solving the
Dirichlet problem for elliptic partial differential equations in complicated
domains. The inequalities used in the convergence proof of the SAP for PDEs
rely on the elliptic nature of the systems.

\begin{figure}[th]
  \epsfxsize=0.5\textwidth
  \centerline{\epspdffile{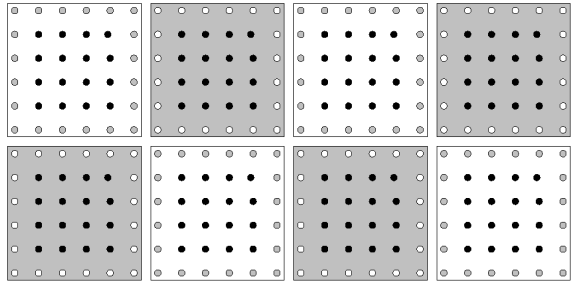}}
  \vskip-6ex
  \caption[Domain decomposition]{SAP domain decomposition.}
  \label{fig:luscher-domains}
\end{figure}

The basic idea is to decompose the lattice into blocks, and then update each
block with boundary conditions specified by the most recent fields on the
neighbouring blocks, as illustrated in \reffig{fig:luscher-domains}. In the
figure \(6^2\) blocks are shown, but when considering the scaling behaviour of
the algorithm the blocks should be taken to be of fixed physical size.

\subsection{SAP Preconditioner for Linear Solver}

When we invert the Dirac operator on the lattice we modify the (pseudo)fermion
fields in a fixed gauge background. For a nearest-neighbour fermion kernel,
such as the Wilson Dirac operator, the fermion fields in the even blocks are
coupled only to those in other odd blocks and vice versa. We can thus alternate
updates of all the even and all the odd block fields, giving a procedure that
has a large degree of parallelism. Notice that we are not just alternating
single updates but we are alternating complete solutions of the Dirichlet
problem on each set of blocks.

\begin{figure}[th]
  \epsfxsize=0.5\textwidth
  \centerline{\epspdffile{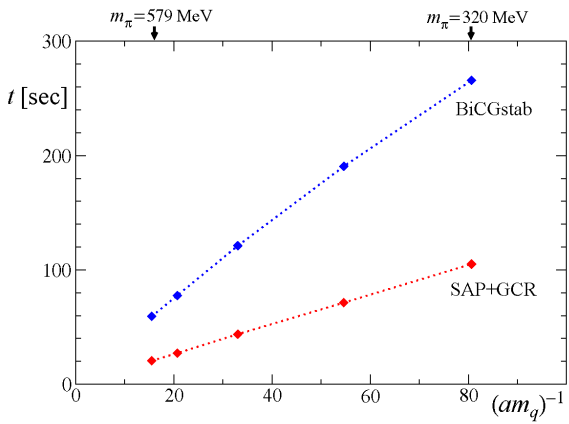}}
  \vskip-6ex
  \caption[SAP performance]{SAP preconditioner performance. The data is for a
  \(48^3\times24\) lattice with \(a=0.10\)~fm, and quark masses between \(0.2\)
  and \(0.7\) of the strange quark mass. The timings were obtained on 8 nodes
  (16 processors) of a PC cluster with a block size of \(6^2\times4^2\) and a
  residual of \(10^{-8}\) relative to the source.}
  \label{fig:luscher-perf}
\end{figure}

This procedure converges to the correct solution, but the rate of convergence
is slow; so instead of using SAP as a solver we may use it as a preconditioner
for a {\Krylov} space solver (such as GCR or FGMRES). For this purpose accurate
block solves are not required, as the preconditioner need only be an
approximation to the inverse, and only a few Schwarz cycles are required. This
method seems to lead to a significant speed-up by a factor of 2--3 over a range
of quark masses, as illustrated in \reffig{fig:luscher-perf}. The algorithm
parallelises easily, especially on coarse-grained architectures such as PC
clusters.

\subsection{Dynamical Fermion MC algorithm}

A more radical use of the SAP is as a dynamical fermion algorithm itself, in
which we update the gauge fields in the blocks. 

The pseudofermion term in the action involves the inverse of the lattice Dirac
operator, and is thus manifestly non-local. In order to allow the update of a
subset of the gauge links we make use of the Schur complement factorisation of
the quark determinant: \(\det\Dslash = \det\Dslash_\Omega \det\Dslash_{
\Omega^{*}} \det\left(1 - \Dslash_\Omega^{-1} \Dslash_{\partial\Omega}
\Dslash_{\Omega^{*}}^{-1} \Dslash_{\partial\Omega^{*}} \right)\). We indicate
the grey and white block of \reffig{fig:luscher-domains} by \(\Omega\) and
\(\Omega^{*}\) respectively, and exterior boundaries (indicated by white dots
in the figure) by \(\partial\Omega\) and \(\partial\Omega^{*}\). This
decomposition is easily derived from the UL factorisation \begin{eqnarray*}
\Dslash & \!\defn\! &\left(\begin{array}{cc} \Dslash_\Omega & \Dslash_{\partial
\Omega} \\ \Dslash_{\partial \Omega^{*}} & \Dslash_{\Omega^{*}} \end{array}
\right) = \left( \begin{array}{cc}1 & \Dslash_{\partial\Omega} \Dslash_{
\Omega^{*}}^{-1} \\ 0 & 1\end{array}\right) \times\\ && \qquad\times
\left(\begin{array}{cc} \Dslash_\Omega - \Dslash_{\partial\Omega} \Dslash_{
\Omega^{*}}^{-1} \Dslash_{\partial \Omega^{*}} & 0 \\ \Dslash_{\partial
\Omega^{*}} & \Dslash_{\Omega^{*}}\end{array} \right). \end{eqnarray*} We may
now introduce two pseudofermion fields, \(\phi\) with support on the interior
of the blocks and \(\chi\) with support on the boundary; the associated
pseudofermion part of the action being \(S_\pf = \phi^\dagger \left(
\Dslash_\Omega^{-2} + \Dslash_{\Omega^{*}}^{-2} \right) \phi + \chi^\dagger
R^{-2} \chi\), with \(R\) being the Schur complement operator introduced
above.\footnote{In practice it is useful to use even-odd preconditioning.} The
gauge fields on the ``active'' links within the blocks together with the
\(\phi\) pseudofermions can be equilibrated with \(\chi\) held fixed, and then
the effect of the ``Schur complement'' interaction \(R\) can be incorporated
into a global HMC process. We occasionally shift the blocks to ensure all links
get their turn to be active and thus updated.

According to the preliminary results that have been presented, this scheme
successfully separates the short- and long-distance effects. Within the blocks
the smallest eigenvalues of the Dirac operator are regulated by the inverse
block size, whereas for the force due to the \(\chi\) pseudofermions is small
and only weakly mass dependent in the interior of the blocks; this allows
larger step sizes for the global HMC process. This looks very promising, and it
will be interesting to see how the costs scale for very small quark masses. So
far the procedure has been applied to Wilson fermions, and it might be painful
to generalise it for those fatter fermion actions for which the fraction of
``active'' links becomes smaller.

\section{Algorithms for Fractional Multiplets}
\label{sec:fractional-multiplets}

Previously in \refsec{sec:locality} we discussed the validity of using a
fractional power of the fermion determinant in order to circumvent the doubling
problem for staggered fermions. In this section we shall ignore such matters of
principal and consider the available repertoire of practical dynamical fermion
algorithms available for this case.

\subsection[R Algorithm]{{\HMDR} Algorithm} \label{sec:hmdr}

To date, all large-scale computations with two or three tastes of staggered
quarks have used the \HMDR~algorithm~\cite{gottlieb87a}. This is an inexact
algorithm, in the sense that it is a Markov process that converges to an
equilibrium distribution that only approximates the desired one. It is based on
the alternation of MD trajectories and Gaussian momentum refreshment, but with
no accept/reject step to correct for non-zero step size errors. By using a
clever combination of non-reversibility and area non-preservation in the MD
integrator, the \HMDR~algorithm converges to a distribution with errors of
\(O(\dt^2)\) in the parameters of the action. As already noted in
\refsec{sec:battlefield} these \(O(\dt^2)\) errors are volume independent.

It is important to note that this result is an asymptotic expansion in \(\dt\),
so there are also errors of the form \(e^{-1/m\,\dt}\) which are only small if
the quantity \(m\,\dt\ll1\) where \(m\) is some suitable scale
parameter. Indeed, we may expect these non-analytic errors to become dominant
at the same \(\dt\) for which the MD becomes
unstable~(\refsec{sec:instability}); we should not expect to be able to use a
larger step size than for the corresponding exact HMC algorithms, except that
instead of getting a tiny acceptance rate we would just quietly get an
incorrect answer. The observation that the estimate of the errors in the
equilibrium distribution is an asymptotic expansion\footnote{ C.f.,
perturbation theory is also only asymptotic (or worse), as is the
``improvement'' expansion in the cut-off \(a\). Indeed, we may reasonably
expect scaling for more highly-improved theories to break down at smaller
values of \(a\).} tells us that we have no reason to expect the results for
large \(\dt\) to just correspond to a renormalisation of the parameters, as was
originally suggested for the related Langevin algorithm.

\subsection{PHMC and RHMC} \label{sec:RHMC}

Relatively recently the Polynomial~\cite{forcrand96a,jansen97a} and
Rational~\cite{kennedy98a,Clark:2003na,Clark:2004cp} Hybrid Monte Carlo
algorithms have been developed. These permit the simulation of fractional
powers of the fermion determinant without any concomitant systematic errors in
the equilibrium distribution.

The basic idea is to use a polynomial or rational function approximation to the
power occurring in the action. Recall that in the context of the HMC algorithm
an approximate action suffices for MD evolution, and an accurate action is only
needed for the acceptance test. This permits us to use a relatively cheap
approximation for each MD step, and a more expensive approximation for the
infrequent acceptance tests.

The fermion kernel \(\M\) on the lattice is a large sparse matrix, and we wish
to use its inverse square root, or more generally some continuous function
\(f(\M)\) of it. We define such functions on Hermitian matrices by
diagonalisation, if \(\M = U\D U^{-1}\) where \(\D\) is diagonal then \(f(\M) =
f(U\D U^{-1}) \defn U f(\D)U^{-1}\). The key property of polynomial and
rational functions is that they do not require the diagonalisation to be
carried out explicitly, since \(\alpha \M^m + \beta \M^n = U(\alpha\D^m +
\beta\D^n) U^{-1}\) and \(\M^{-1} = U\D^{-1}U^{-1}\).

\subsection[Chebyshev Approximation]{{\Chebyshev} Approximation}

\begin{figure}[th]
  \epsfxsize=0.45\textwidth
  \centerline{\epspdffile{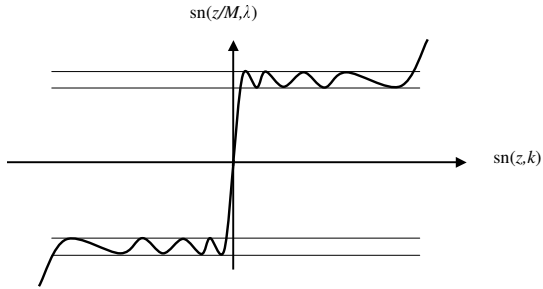}}
  \vskip-6ex
  \caption[Elliptic 2D]{The optimal {\Chebyshev} approximation to the function
  \(\sgn x\) for \(\varepsilon\leq|x|\leq1\), illustrating the characteristic
  alternating extrema of the error. In this case the coefficients of the
  optimal rational approximation were found in closed form by {\Zolotarev} in
  terms of {\Jacobi} elliptic functions, as indicated by the labels on the
  axes.}
  \label{fig:elliptic-2d}
\end{figure}

The theory of optimal \(L_\infty\) (\Chebyshev) approximation over a compact
interval is well understood: \Chebyshev's theorem tells us that there is a
unique optimal approximation characterised by having alternating error maxima
of equal magnitude (c.f., \reffig{fig:elliptic-2d}). In general we can
determine the coefficients of the optimal approximation numerically using
\Remez' algorithm, but for the interesting cases of \(\sgn(x)\) and
\(x^{\pm\half}\) they are have been calculated in closed form in terms of
elliptic functions by \Zolotarev.

It is not necessary to use the optimal approximation, but doing so leads to
surprisingly small errors; for example, if we consider rational approximations
to \(\sgn(x)\) of degree \((21,20)\) then the optimal {\Chebyshev/\Zolotarev}
approximation has a maximum error of \(4.38\times10^{-3}\) over \(10^{-6}\leq
|x|\leq1\), whereas the Higham rational approximation of the same order,
\(\tanh(20 \tanh^{-1}x)\), clearly has a maximum error of about~1.

It is worth emphasising the r\^ole played by {\Chebyshev} polynomials in the
theory of optimal approximation. These polynomials are defined by \(T_n(x) =
\cos(n \cos^{-1}x)\), and have exactly \(n+1\) alternating extrema of unit
magnitude over the interval \(-1\leq x\leq1\). This means that \(x^n - 2^{1-n}
T_n(x)\) is the best {\Chebyshev} approximation to \(x^n\) of degree
\(n-2\).\footnote{{\Chebyshev} polynomials also form an orthonormal basis for
\(L^2[-1,1]\) with respect to the weight \(1/\sqrt{1+x^2}\), up to a trivial
factor in the normalisation of \(T_0(x)\). Nevertheless, truncated {\Chebyshev}
expansions are not optimal in general, even for polynomials.}

There are some advantages of rational over polynomial approximations. In most
cases the error for approximations of a given degree are much smaller for
rational approximations; for example for the \(L_\infty[-1,1]\) error for
\(|x|\) is proportional to \(e^{n/\ln\varepsilon}\) for the best rational
approximation of degree \(n\) whereas the best polynomial approximation has an
error that only falls as \(1/n\). The rational approximation's exponential
convergence means that it can be made exact to machine precision in a
numerically stable and cheap manner. The obvious advantage of a polynomial
approximation is that it does not require the numerical solution of systems of
linear equations, but bear in mind that the polynomial approximation to \(1/x\)
corresponds to computing the matrix inverse using the {\Jacobi} method, as
compared with the more efficient {\Krylov} solvers that are used in the
computation of rational approximations. If the rational functions are expressed
as a partial fraction expansion, and a multi-shift solver~\cite{Frommer:1995ik}
is used to compute all the necessary inverses simultaneously in the same
{\Krylov} space, then rational approximations are very competitive.

\section{Instability of Symplectic Integrators} \label{sec:instability}

In order to understand some of recent techniques for decreasing the cost of
dynamical fermion HMC computations by increasing the integration step size we
need to briefly review some earlier results on the mechanisms which limit these
step sizes~\cite{Joo:2000dh}.

\begin{figure}[th]
  \epsfxsize=0.46\textwidth
  \centerline{\epspdffile{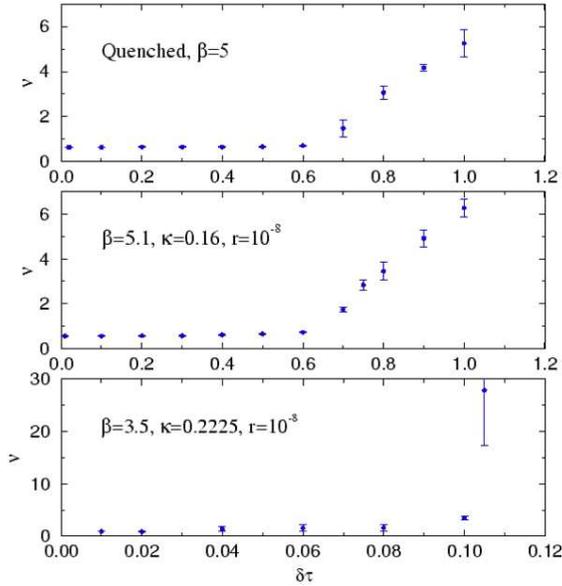}}
  \vskip-6ex
  \caption[Instability]{The {\Liapunov} exponent \(\nu\) for the leapfrog
  integration sheme for quenched QCD (top), QCD with heavy dynamical fermions
  (middle), and QCD with light dynamical fermions (bottom). The exponents were
  measured as a function of the step size \(\dt\) by measuring the violation of
  reversibility caused by rounding errors during a single MD trajectory. Note
  the difference in the scales of the bottom graph and the other two.}
  \label{fig:instability}
\end{figure}

In order to satisfy detailed balance and thus be an exact algorithm we need to
use an area-preserving reversible MD integration scheme within the HMC
algorithm. Fortunately symmetric symplectic integrators, of which the familiar
leapfrog scheme is the lowest-order example, have these desirable
properties. They are only exactly reversible up to floating-point rounding
errors, and these errors become important when they are exponentially amplified
by the chaotic nature of the underlying equations of
motion~\cite{jansen96a,kennedy96c}. This is illustrated in
\reffig{fig:instability}, where the {\Liapunov} exponent \(\nu\) that
characterizes this chaotic amplification is plotted as a function of the
integration step size~\(\dt\). Notice that \(\nu>0\;\forall\dt\) in all three
graphs; this corresponds to the fact the underlying continuous time equations
of motion are indeed chaotic. This interpretation is supported by the fact that
the \(\nu\) seems to have a constant value independent of \(\dt\). When \(\dt\)
exceeds some critical value \(\dt_c\) then \(\nu\propto\dt\); this corresponds
to an instability of the leapfrog integrator, a phenomenon which occurs for the
leapfrog scheme even for a single harmonic oscillator. \(\dt_c\) depends on
quark mass: for the quenched theory or the case of heavy fermions the
instability occurs at values of \(\dt\) so large that the acceptance rate is
already essentially zero, since \(\dH(\dt_c)\gg1\); with light fermions the
instability sets in for much smaller values of \(\dt\), and is what prevents
larger step sizes from being used. Indeed, this is the reason why the use of
higher-order symmetric symplectic integrators is not useful; they increase the
exponent \(n\) characterising the growth of the bulk contribution, \(dH \propto
\dt^n\), but yet again this is only an asymptotic expansion and the
contributions of the form \(e^{-1/\nu\,\dt}\) dominate when \(\dt>\dt_c\).

\subsection{Multipseudofermions}

In order to include the dynamical effects of fermions we need to evaluate a
functional integral including the fermionic determinant \(\det\M\). The usual
method of doing so is to write this as a bosonic functional integral over a
pseudofermion field with kernel \(\M^{-1}\), using the identity \(\det\M
\propto \int d\phi^{*} d\phi\, \exp\left( -\phi^{*} \M^{-1}\phi\right)\). We
then alternate Markov steps of (i)~selecting a pseudofermion field from a
Gaussian heatbath with the gauge fields \(U\) held fixed, and of (ii)~updating
the gauge fields, e.g., by using an HMC algorithm, while keeping the
pseudofermion fields fixed.

This is a perfectly valid procedure, but we are introducing extra noise into
the system by only using a single pseudofermion field to sample the functional
integral. This noise manifests itself as fluctuations in the force exerted by
the pseudofermions on the gauge fields; this increases the maximum fermion
force, which may trigger the integrator instability; and this in turn requires
us to decrease the integration step size.

A less noisy estimate is to use the identity \(\det\M=[\det\M^{\rational1n}]^n
\) to introduce \(n\) \emph{multipseudofermion} fields, \[\det\M \propto
\prod_{j=1}^n \int d\phi_j^{*} d\phi_j\, \exp\left( -\phi_j^{*}
\M^{-\rational1n}\phi_j\right);\] this should increase the value of \(\dt_c\).

\subsubsection{Some Comments on Pseudofermions}

In the limit of an infinite number of multipseudofermions the force on the
gauge fields is computed ``exactly,'' but this is not the same as computing the
force due to \(\det\M\), because the pseudofermions are updated at beginning of
a gauge field trajectory, and do not represent \(\det\M\) except at the
beginning of the trajectory.\footnote{Except for Langevin type, or Generalised
HMC (Kramers) algorithms.}

It is not at all clear that this is in any way a problem, however. Alternating
gauge field trajectories with pseudofermion heatbath updates is a valid Markov
chain with the correct equilibrium distribution. Furthermore, the MD ensures
that the gauge fields evolve along a trajectory in the fixed pseudofermion
background, and this is not the same trajectory as they would take under the
``instantaneous'' force due to \(\det\M\). There appears to be no reason for
the pseudofermion force along a trajectory to be larger for one sort of
trajectory that for the other: they are just different Markov chains with the
same fixed point distribution.

\subsubsection{Reduction of Maximum Force}

The first method of implementing multipseudofermions was introduced by
Hasenbusch~\cite{Hasenbusch:2001ne}. For the case of the Wilson fermion action
\(\M=1-\kappa H\) he introduced the quantity \(\M' = 1-\kappa'H\), and used the
identity \(\M = \M'(\M'^{-1}\M)\) to write the fermion determinant as \(\det\M
= \det\M' \det(\M'^{-1}\M)\). He then used a separate pseudofermion for each
determinant, and tuned \(\kappa'\) to minimise the cost.

This method easily generalises to more than two pseudofermions, and to more
complicated actions such as the Wilson-clover action.

\subsubsection{RHMC Force Reduction}

Another way of implementing multipseudofermions is to use the RHMC technique
(\refsec{sec:RHMC}) to implement \(n\)th root of the fermion kernel
\cite{Clark:2004cq}. Following the usual procedure for RHMC we use a partial
fraction expansion of the optimal rational approximation for \(n\)th root, and
use a multishift solver to compute all the terms in the same {\Krylov} space.

The advantage of this approach is the no tuning is required, as the condition
number \(\kappa(\M)\) is divided equally among all the multipseudofermions. If
we make the na{\"\i}ve assumption that the cost is proportional to condition
number of the matrices we need to invert, then the condition number for each
multipseudofermion kernel is \(\kappa(r(\M)) = \kappa(\M)^{1/n}\), so the total
fermionic force is reduced by factor \(n\kappa(M)^{(1/n)-1}\). If this permits
us to increase the step size by the same factor then the optimal value \(n_\opt
\approx\ln\kappa(M)\) minimises the cost, with an optimal cost reduction of
\(e\ln\kappa/\kappa\).

Of course, this argument is too na{\"\i}ve, we cannot reduce force by an
arbitrarily large factor just by increasing the number of multipseudofermion
fields~\(n\). At some point the fluctuations in the force coming from the
stochastic estimate of the fermion determinant are no longer larger than the
force itself, at which point increasing \(n\) much further will not help.

\section[Reducing dH Fluctuations]{Reducing \(\dH\) Fluctuations}
\label{sec:integrator-tricks}

\subsection{New Integrator Tricks}

\begin{figure}[t]
  \epsfxsize=0.25\textwidth
  \centerline{\epspdffile{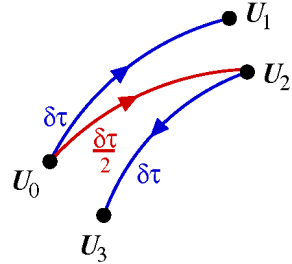}}
  \vskip-7ex
  \caption[Luscher wiggle]{Trajectories used in the \Luscher-Sommer trick.}
  \label{fig:luscher-wiggle}
\end{figure}

An interesting ``replay'' trick for reducing fluctuations in \(\dH\) along an
HMC MD trajectory has been suggested by {\Luscher} and
Sommer~\cite{Luscher:2004}. Consider the situation shown in
\reffig{fig:luscher-wiggle}. The system starts at point \((U_0,\pi_0)\) in
phase space, and the energy change along the trajectory to \((U_1,\pi_1)\) with
step size \(\dt\) is \(\dH_{0\to1}\). If \(\dH_{0\to1}\) is small, say
\(|\dH_{0\to1}|\leq1\), then we accept the new configuration with probability
\(\min\left(1, e^{-\dH_{0\to1}}\right)\) or else keep the old configuration
\(U_0\); but if \(|\dH_{0\to1}| > 1\) we construct a more accurate trajectory
from \((U_0, \pi_0) \to (U_2,\pi_2)\) with step size \(\dt/2\). In order to
ensure reversibility we need to make sure that this same reduced step size
would be used if we started at \((U_2,-\pi_2)\), so we must verify that the
fictitious energy change along the trajectory from \((U_2,-\pi_2) \to
(U_3,-\pi_3)\) with step size \(\dt\) has energy change \(|\dH_{2\to3}| > 1\).
If this is the case we accept \(U_2\) with probability \(\min\left(1,
e^{-\dH_{0\to2}} \right)\) or else keep \(U_0\), and everything is
area-preserving and reversible. If \(|\dH_{2\to3}|\leq1\) then we would not try
reducing the step size in the reverse direction, so we fall back to accepting
\(U_0\). By considering the cases where the energy changes \(\dH_{0\to1}\),
\(\dH_{2\to3}\), and \(\dH_{0\to2}\) are negative, between 0 and 1, and \(>1\)
separately detailed balance is easily verified.

This procedure is very useful if the system occasionally hits a large energy
fluctuation, for example if it comes across an integrator instability. If such
instabilities occur frequently then we must ask whether it would not be
preferable just to reduce the step size (say to \(\dt/2\)) for all
trajectories. Another issue to consider is that often these large \(\dH\)
changes occur as the system is approaching equilibrium, and we will be forced
to reduce \(\dt\) anyhow as the system becomes more equilibrated and ``sees''
the large force due to the light fermions.

\subsection{Old Integrator Tricks}

Another interesting trick that is sometimes useful was introduced by Sexton
and Weingarten~\cite{sexton92a} over a decade ago. Their idea is to split MD
Hamiltonian into two parts, usually corresponding to the boson and fermion
parts of the action, and to construct a symmetric symplectic integrator with
larger steps for more expensive (fermion) part by a clever application of the
Baker--Campbell--Hausdorff formula. This helps if the step size is limited by
cheaper (boson) part, but unfortunately becomes less useful as \(m\to0\) when
the step size is limited by the integrator instabilities induced by the fermion
force. Nevertheless, it is a useful technique when the large contributions to
the force are not also the most expensive.

\section{Dynamical Chiral Fermions} \label{sec:chiral}

We now turn to one of the most important developments of recent years: the
formulation of on-shell chirally symmetric fermions on the lattice. Our goal is
to show that the various techniques developed for this purpose, such as the
domain wall or overlap formulations, are equivalent, in the sense that they are
just different approximations to the same on-shell chiral lattice theory. There
are many variants of the methods, but they all seem to correspond to choosing
(explicitly or implicitly) a rational approximation to the \(\sgn\) function,
and of computing the inverse of the corresponding approximate Neuberger
operator.

Let us fix our notation and conventions: we shall work in Euclidean space with
Hermitian \(\gamma\) matrices; and we shall take all Dirac operators to be
\(\gamma_5\) Hermitian, \(\Dslash^\dagger = \gamma_5\Dslash\gamma_5\).

\subsection{On-shell chiral symmetry}

Although it was not the way the subject developed historically (a matter that
we will not even attempt to review here), an elegant logical approach
\cite{Luscher:1998pq} is to ask if it is possible to have chiral symmetry on
the lattice without doublers if we only insist that the symmetry holds on
shell. Such a transformation would have to be of the form \[\psi \to
e^{\alpha\gamma_5(1-a\Dslash/2)} \psi,\quad \bar\psi \to\bar\psi e^{\alpha
(1-a\Dslash/2)\gamma_5},\] which becomes a chiral transformation on-shell where
\(\Dslash\psi=0\). Note that we are free to specify the chiral transformation
properties of \(\psi\) and \(\bar\psi\) independently. For this transformation
to be a symmetry the Dirac operator must be invariant \[\Dslash \to e^{\alpha
(1-a\Dslash/2)\gamma_5} \Dslash e^{\alpha\gamma_5(1-a\Dslash/2)} = \Dslash.\]
For a small transformation, \(\alpha\ll1\), this implies that \((1-\half
a\Dslash)\gamma_5 \Dslash + \Dslash \gamma_5(1-\half a\Dslash) = 0\), which is
just the Ginsparg--Wilson (GW) relation \(\gamma_5\Dslash + \Dslash\gamma_5 =
a\Dslash\gamma_5\Dslash\).

\subsection{Neuberger's Operator}

We can find a solution of the Ginsparg--Wilson relation as follows: let the
lattice Dirac operator be of the form \(a\Dslash = 1 + \gamma_5\hat\gamma_5\),
where \(\hat\gamma_5^\dagger = \hat\gamma_5\), thus \(a\Dslash^\dagger = 1 +
\hat\gamma_5\gamma_5 = \gamma_5a\Dslash\gamma_5\). This satisfies the
Ginsparg--Wilson relation if \(\hat\gamma_5^2=1\), and it has the correct
continuum limit \(\Dslash\to\dslash\) if \(\hat\gamma_5 = \gamma_5 (a\dslash-1)
+ O(a^2)\). Both of these conditions are satisfied by the choice \[\hat
\gamma_5 = \frac{\gamma_5 (\DWslash-1)} {\sqrt{(\DWslash-1) (\DWslash-
1)^\dagger}} = \sgn[\gamma_5(\DWslash-1)].\] There are many other possible
solutions, but these essentially all correspond to just using a different Dirac
operator within Neuberger's operator.

We should emphasise that we are only considering vector-like theories with a
chiral symmetry here, chiral theories with unpaired Weyl fermions can be
discretised on the lattice, but getting the phase of the fermion measure
correct is critical. Simulating such theories is a much harder problem than the
already challenging computations required for the case considered here.

All numerical techniques for calculating with on-shell chiral fermions may be
viewed as different ways of approximating Neuberger's operator. It is
relatively straightforward if somewhat expensive to apply the operator, for
instance by using a rational or polynomial approximation to the \(\sgn\)
function, but it is more challenging to apply its inverse. The difficulty is
that Neuberger's operator is constructed out of two non-commuting operators,
\(\gamma_5\) and \(\hat\gamma_5\), so the inverse of a rational approximation
is not just the reciprocal rational function.

There are various ways the inverse may be computed: the most direct approach is
by a nested {\Krylov} space approach, where for every outer iteration a full
inner inversion is required to apply the Neuberger operator. 

The obvious problems with this approach are (i)~the information built up in the
inner {\Krylov} space is not made use of in the outer {\Krylov} space; (ii)~the
accuracy (residual) required for the inner solves in order to reach a specified
accuracy for the outer solve is not immediately obvious. Quite a lot of work
has gone into developing these techniques, including recent studies of various
preconditioners and the use of ``flexible inverters'' for which the
approximation used for the preconditioner may vary during the outer
inversion~\cite{Krieg:2004xg,vandenEshof:2002ms,Arnold:2003sx,Cundy:2004pz}.

\subsection{Into Five Dimensions}

An alternative approach to computing the inverse of the Neuberger operator is
to express it as a single inverse of a five dimensional operator. This may done
is by linearising a continued fraction representation of a rational
approximation to the Neuberger operator~\cite{Neuberger:1998my,Borici:2002a} by
the introduction of a set of \(n_5\) auxiliary spinor fields; these can be
introduced either as constraints or as new dynamical variables. From this
viewpoint the new ``fifth'' dimension is just an means of approximating the
\(\sgn\) function, and the gauge fields are still only four dimensional. 

This is similar to the domain wall formulation~\cite{Furman:1994ky}, and it is
possible to translate domain wall fermions into an equivalent four-dimensional
approximation to the Neuberger operator~\cite{Borici:1999zw,Borici:1999da,%
Borici:2004pn,Edwards:2000qv}. To do this we first separate the left- and
right-handed parts of the domain wall kernel and cyclically shift the latter by
one site in the fifth dimension. In the domain wall formalism the
four-dimensional fermion fields have their left-handed components on one wall
and their right-handed components on the other wall, and the shift moves them
both to the same wall, as well as reducing the kernel to a form in which we can
identify an explicit fifth-dimensional transfer matrix~\(T\). In this form we
can integrate out the bulk fermion fields, cancelling the Jacobian with the
contribution from the five-dimensional pseudofermion fields, and construct an
effective four dimensional Hamiltonian \(H\defn\frac{T-1}{T+1}\), which is the
Euclidean form of a Cayley transform of the transfer matrix. This gives a
Higham approximation \(\sgn x \approx \tanh(n_5 \tanh^{-1} x)\) to the
Neuberger operator \(\Dslash = \half(1 + \gamma_5\sgn H)\).

Bori\c{c}i also introduced the ``truncated overlap'' for which \(H = \gamma_5
\DWslash\), and Chiu has shown how to modify the domain wall formalism to give
a {\Zolotarev} approximation in the equivalent effective four-dimensional
theory \cite{Chiu:2002ir,Chiu:2002kj,Chiu:2003ir}. At this meeting Brower,
Originos, \& Neff \cite{Brower:2004xi} introduced ``M\"obius fermions,'' which
interpolate between domain wall and truncated overlap fermions.

\subsection{Overlap Algorithms}

We see that there are many formulations of overlap fermions, all of which
satisfy the GW relation, and which are equivalent in continuum limit. The
principal differences between them are that they use different lattice Dirac
operators within the \(\sgn\) function, and different approximations to the
\(\sgn\) function. This is presumably also true for perfect actions. The choice
between the formulations is essentially just a trade-off between computational
speed and the amount of chiral symmetry breaking.

\subsection{Dynamical Overlap}

To date, most applications of on-shell chiral fermions have been for valence
computations, but of course we need to introduce dynamical (sea) overlap
fermions too in order to be able to obtain reliable chiral results from full
QCD. The first studies of dynamical overlap were carried out in the Schwinger
model~\cite{Edwards:1998yw,Bode:1999dd} several years ago. Large-scale
dynamical domain-wall QCD computations have been carried out by the RBC
collaboration~\cite{Fleming:1999xe,Chen:1998xw,Vranas:1998vm}. Recently Fodor,
Katz, \& Szab\'o~\cite{Fodor:2003bh,Fodor:2004wx} and Cundy, Krieg, Frommer,
Lippert, \& Schilling~\cite{Cundy:2004xf} have studied dynamical overlap
fermions in QCD on ``ridiculously small lattices.''

Apart from the issues in common with valence overlap computations that we have
discussed previously, HMC computations involving overlap fermions require the
computation of the MD force. In particular, there is an issue concerning what
to do when one of the eigenvalues of \(\gamma_5\DWslash\) passes through zero,
which for continuous time MD evolution with an exact Neuberger operator would
lead to a singular force.

If a polynomial or rational approximation to the \(\sgn\) function is used then
this is readily differentiable and leads to a computable force term, so one
method of handling the singular force is just to observe that there is no
discontinuity in the approximations to the \(\sgn\) function. This is
effectively the method used in the dynamical domain wall computations, where
the Higham approximation is implicitly used and which, for practicable values
of \(n_5\), smoothes the discontinuity enough to keep the force under
control. Of course, this is just a reflection of the fact that chiral symmetry
is not very well approximated. Another approach \cite{Fodor:2003bh} is to
modify the leapfrog integrator to reflect or refract at a zero crossing in such
a way that the MD process is still exactly reversible and area preserving. It
will be interesting which, if any, of these approaches avoids the HMC
acceptance rate becoming small due to zero crossings.

We also note the suggestion~\cite{Bode:1999dd} of working in the chiral sector
with no zero modes, and explicitly reweighting the configurations with
\(m^{2Q}\) where \(Q\) is the topological charge.

\section{Conclusions}

We are at an interesting point in the development of algorithms for dynamical
fermion computations in QCD. On the one hand it is now generally accepted that
quenching leads to significant systematic errors, and that to proceed towards
reliable higher accuracy calculations we need to include dynamical fermions one
way or another. On the other hand, there is much debate about whether the
systematic errors of two or three flavour staggered quark simulations are under
control. Recent progress with dynamical {\qcdtm} and GW fermions also show that
the prospects for carrying out large-scale dynamical computations with light
quarks, with all sources of systematic errors under control, and even with good
chiral symmetry properties should be expected with the current or next
generation of computers.

\section*{Acknowledgments}

I would like to thank Martin {\Luscher} and Urs Wenger for making helpful
comments on draft versions of this article.

\raggedright

\end{document}